# Resolving the discrepancy between MOKE measurements at 1550-nm wavelength on Kagome Metal CsV$_3$Sb$_5$

Jingyuan Wang[1], Camron Farhang[1], Brenden R. Ortiz[2], Stephen D. Wilson[2] and Jing Xia[1]

[1]Department of Physics and Astronomy, University of California, Irvine, CA 92697, USA
[2]Materials Department, University of California, Santa Barbara, Santa Barbara, CA 93106, USA

Kagome metals AV$_3$Sb$_5$ (A = K, Cs, Rb) provide a rich platform for intertwined orders such as the charge density wave (CDW) and a chiral order with time-reversal symmetry breaking (TRSB). While early reports of large optical polarization rotations have been interpreted as the magneto-optic Kerr effect (MOKE) and as evidence for TRSB, recent dedicated optical rotation and MOKE experiments have clarified that this large optical rotation originates instead from an unconventional specular rotation. Yet a critical discrepancy remains regarding the possible existence of a true spontaneous MOKE signal: in experiments performed after training with modest magnetic fields of up to 0.3 Tesla, no MOKE signal was detected above the noise floor of 30 nano-radians, while micro-radian-level signals were found in an experiment using higher training fields. This raises an intriguing possibility of different zero-field ground states with opposite time-reversal symmetry properties, because of different magnetic histories. To unambiguously determine whether a training-field-dependent spontaneous MOKE signal exists in CsV$_3$Sb$_5$, we conduct comprehensive MOKE measurements with two Sagnac interferometer setups capable of both low and high training fields of up to 9 Tesla, and perform careful analyses of contributions of signals from various optical components. We conclude that there is no observable spontaneous MOKE signal, hence no optical evidence for TRSB, regardless of the magnitude of training fields and the speed of temperature ramping.

## I. INTRODUCTION

Kagome metals featuring elaborate lattice structures and rich varieties of quantum phenomena have stimulated intense experimental efforts to uncover novel phases of matter where strong correlation and topological orders intertwine. The recently discovered quasi-two-dimensional Kagome compound family AV$_3$Sb$_5$ (A=K, Rb and Cs) [1,2] is of great current interest due to its exhibition of charge density wave (CDW) [3–7], pressure-tunable superconductivity [2,8–11], and possibly time reversal symmetry breaking (TRSB) [12–14]. In particular, a TRSB order parameter in the CDW state has been reported in some of the muon-spin relaxation (μSR) [12–14], chiral transport [15] and optical polarization rotation [16,17] experiments, and may have connections with the long-sought loop currents [18], originally proposed for cuprate superconductors. AV$_3$Sb$_5$'s rich phase diagrams originate from the ideal Kagome network governed by layers of vanadium and antimony intercalated by alkali metal ions that crystallize in the p6/mmm group [1,2]: tight-binding models of such Kagome lattices have revealed a fascinating electronic band structure containing Dirac cones, flat bands, and Van Hove singularities [2,19], making AV$_3$Sb$_5$ an interesting platform with intertwined [20] electronic instabilities for exotic correlated phases.

A hint for such an exotic phase in AV$_3$Sb$_5$ came from intriguing and contradicting magneto-optic Kerr effect (MOKE) experiments [16,17,21–23] that defy a unified explanation with existing theories including the loop currents model [18]. A spontaneous (at zero magnetic field) Kerr rotation $\theta_K$, which arises from the difference between refraction indices ($n + ik$) of time-reversed circularly polarized light beams, is direct evidence for TRSB. The majority of MOKE experiments have been conducted on the Cs compound CsV$_3$Sb$_5$ and have yielded disparate spontaneous $\theta_K$ values below the CDW transition temperature $T_{CDW} \sim 94\ K$, ranging from $450\ \mu rad$ [17] to $50\ \mu rad$ [16] at $800\ nm$ wavelength, to $2\ \mu rad$ [23] at $1550\ nm$ wavelength and less than $0.03\ \mu rad$ [21,22] at $1550\ nm$ wavelength. This astounding discrepancy needs to be addressed immediately.

The difference between $1550\ nm$ and $800\ nm$ experiments has recently been resolved [22]. The above $800\ nm$ experiments [16,17] were performed at zero magnetic field using the optical rotation technique. In such an experiment, the (total) polarization rotation $\theta_T$ of a linearly polarized light is measured upon reflection from the sample as a function of angle $\alpha$ between light polarization and the sample's principal axis. $\theta_T$ is then fitted to equation $\theta_T = \theta_C + \theta_P \sin(2\alpha)$, where $\theta_P$ is the angle-dependent polarization rotation induced by rotational-symmetry-breaking, and $\theta_C$ is the angle-independent (isotropic) component. $\theta_C$ is usually assumed to be the MOKE signal ($\theta_K$), which should flip sign with opposite magnetic fields. This hypothesis, however, was rejected recently by field-dependent measurements at $1550\ nm$ wavelength [22], which has revealed a large $\theta_C$ signal in CsV$_3$Sb$_5$ that doesn't reverse sign with oppositely oriented magnetic fields, indicating that $\theta_C$ is not directly related to TRSB. In fact, the sub-milli-radian size of $\theta_C$ is simply too big to be associated to TRSB in CsV$_3$Sb$_5$: with the sub-Gauss level internal magnetic field measured in (μSR) experiments [12–14], the spontaneous MOKE signal $\theta_K$, if existent, is expected to be at the level of nanoradians to microradians [24,25].

The existence of this spontaneous MOKE signal has been tested recently [21–23] with a loopless fiber-optic Sagnac interferometer [26], which by design is sensitive only to TRSB effects with the required nanoradian-level MOKE sensitivity. Yet puzzling discrepancies were found between these Sagnac interferometer experiments, yielding spontaneous Kerr signals $\theta_K$ ranging from 2 $\mu rad$ [23] to less than 0.03 $\mu rad$ [21,22]. Although performed at the same 1550 $nm$ wavelength, these measurements have important differences regarding magnetic training fields, temperature ramping rates, and detailed optical configurations. In a typical Sagnac experiment aimed at testing spontaneous MOKE signal, the sample is first cooled down from high temperature with a magnetic training field to align potential chiral domains. The field is then removed at low temperature and the MOKE measurement is performed during zero-magnetic-field warmup. In two Sagnac experiments [21,22] performed after training with modest magnetic fields ranging from 0.01 Tesla to 0.3 Tesla, no MOKE signal was detected above the noise floor of 0.03 $\mu rad$. In sharp contrast, MOKE signals of 2 $\mu rad$ were found in another experiment [23] using much higher training fields up to 10 Tesla.

Does the above discrepancy between Sagnac measurements indicate two distinct zero-field ground states in CsV$_3$Sb$_5$ with opposite time-reversal symmetry properties, depending on the magnetic history? Here we conduct comprehensive MOKE measurements at both low and high training fields of up to 9 Tesla to unambiguously determine whether a training-field-dependent spontaneous MOKE signal exists in CsV$_3$Sb$_5$. In addition, as a delicate optical technique [26], it is important to align a Sagnac interferometer properly to eliminate false signals from subtle details in the experimental configurations. Therefore, in this work we have used two Sagnac instruments that are similar to those used in [21,22] and in [23] respectively, and we perform careful analyses of contributions of signals from various optical components.

We report the repeatable and condition-independent absence of spontaneous MOKE in the same CsV$_3$Sb$_5$ sample in both Sagnac instruments with 0.1 $\mu rad$ uncertainty, and conclude that there is no observable optical evidence for TRSB, regardless of the magnitude of training fields and the speed of temperature ramping. In addition, we demonstrate that subtle artifacts in the "cryo-optics" setup (Fig.1(a)), if not treated properly, can yield incorrect temperature-dependent measurements of $\theta_K$. This leads to a wrong sign of $\theta_K$ that was recently acknowledged in version 2 of [23], and may lead to a false spontaneous $\theta_K$ signal.

## II. SAGNAC MOKE MICROSCOPES FOR HIGH AND LOW MAGNETIC FIELDS

Before presenting our results, it's worthwhile to review some important details in the prior 1550 $nm$ Sagnac experiments [21–23]. They are all based on zero-loop Sagnac interferometry [26] that measures directly the optical phase difference between counter-propagating circularly polarized light beams. It is hence only sensitive to TRSB effects, and can achieve MOKE ($\theta_K$) sensitivities of $\approx$ 0.01 $\mu rad$ [24–27].

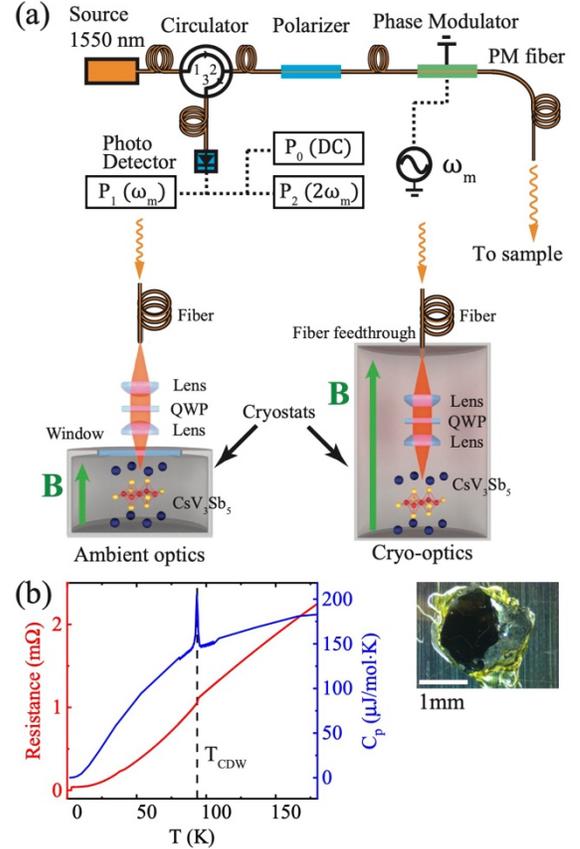

FIG. 1. Experimental configurations and CsV$_3$Sb$_5$ sample. (a) A Sagnac interferometer is connected to cryostats with either "ambient optics" (used in [21]) or "cryo-optics" (used in [23]). The "ambient optics" allow correct background removal in the presence of magnetic fields $B$, at the expense of limited size of $B$. (b) The CsV$_3$Sb$_5$ crystal used in this study with $T_{CDW} \sim 94\ K$ marked by a kink in the resistance, and a sharp peak in the specific heat $C_P$.

A schematic of this interferometer is shown in Fig. 1(a), where time-reversed light beams exiting the polarization maintaining (PM) optical fiber are converted by a quarter-wave plate (QWP) to opposite circular polarizations and interact with the sample during reflection. The returned light beams interfere at the detector and the MOKE signal $\theta_K$ is proportional to the ratio between the first harmonic power $P_1$ and the second harmonic power $P_2$ as explained in details in [26]. In a typical Sagnac experiment, $\theta_K$ is recorded first during cooldown with a magnetic field. The measured

MOKE susceptibility $\theta_K/B$ gives a measure of magnetic susceptibility $\chi$. In the subsequent zero field warmup (ZFW) spontaneous $\theta_K$ can be measured with TRSB domains aligned by the prior "training field" [21,23].

On CsV$_3$Sb$_5$, the experiments performed at Stanford and Irvine [21,22] show no spontaneous $\theta_K$ with an uncertainty of $0.03\ \mu rad$ after field trainings up to $0.03\ T$ and $0.3\ T$ respectively. In contrast, the Kyoto experiment [23] has revealed below $T_{CDW}$ a spontaneous $\theta_K \approx +2\ \mu rad$ ($-2\ \mu rad$) after positive (negative) field trainings with much larger training fields of up to $10\ T$.

This vast difference in the sizes of training magnetic fields has been argued to be the most likely cause for the opposite behavior of spontaneous $\theta_K$ between these Sagnac experiments. Indeed, metamagnetic phases with different magnetic configurations exist in strongly correlated materials such as those observed in the heavy Fermion metal CeAgBi$_2$ [28]. And the CDW states in AV$_3$Sb$_5$ are known to be tunable by magnetic fields [29–32]. Considering the first order nature of the CDW transition, it is plausible that different zero-field ground states can be realized after removing the magnetic fields, yielding different spontaneous $\theta_K$.

Temperature ramping rate is another important factor that can modify the zero-field ground state of a magnetic system. One well-known example is the skyrmion systems, where large cooling rates are crucial to stabilize the metastable skyrmions due to their very short lifetime [33,34]. The Kyoto experiments used a much slower rate compared to Stanford and Irvine, which could potentially induce a different magnetic ground state at zero field that gives spontaneous $\theta_K$.

The technical reason for the very different magnetic fields in these studies [21–23] is related to the optical cryostats used. As illustrated in Fig.1(a), the Stanford and Irvine experiments [21,22] were performed with the "ambient-optics" setup (Fig.1(a) left) with all optics outside the cryostat that was usually used for low field experiments [35,36]. The Kyoto experiment [23] was conducted with a "cryo-optics" setup (Fig. 1(a) right) with the QWP and lens inside the cryostat, that was normally for experiments requiring high fields [37] or extremely low temperatures [24].

At zero magnetic field, in principle there should be no difference in measured $\theta_K$ between the two configurations. However, when a magnetic field is applied, or when there is a remanent field in a superconducting magnet, the optical components will have their own temperature-dependent contributions to $\theta_K$ via the magneto-optic Faraday effect, which need to be removed to obtain accurate readings of $\theta_K$ from the sample. Such calibrations can only be performed properly with "ambient-optics" that are at fixed temperatures. Calibrations for the Irvine instrument have been described in [21,22]. In contrast, due to the long thermalization time of insulators at cryogenic temperatures, the temperatures of "cryo-optics" deviate significantly from sample temperature and are rather non-repeatable between thermal cycles. As a result, the signal subtraction described in [23] with the "cryo-optics" setup becomes problematic. As an example, a wrong sign of $\theta_K$ has recently been acknowledged in version 2 of [23].

In this work we utilize both "cryo-optics" and "ambient-optics" configurations to unambiguously determine whether a training-field-dependent spontaneous MOKE signal exists in CsV$_3$Sb$_5$. This comparative study also allows us to demonstrate that in an applied magnetic field, although the absolute value of $\theta_K/B$ can't be obtained properly with the "cryo-optics" setup, the sharp change $\Delta\theta_K/B$ across $T_{CDW}$ can be determined quite accurately in both setups.

## III. RESULTS

The high quality CsV$_3$Sb$_5$ single crystal (inset in Fig. 1(b)) used in this study is similar to those used in the prior Irvine Sagnac and optical rotation experiments [21,22], and was grown by self-flux method at UC Santa Barbara [2].

Heat capacity $C_P$ and electric resistance are measured in zero magnetic field as functions of temperature and are presented in Fig. 1(b). The first order CDW transition at $T_{CDW}\sim 94\ K$ is clearly characterized by the sharp peak in the heat capacity $C_P$ and a kink in the resistance, both at $T_{CDW}$.

The crystal was then cleaved to expose optically flat areas and was mounted to both optical cryostats using Ge-varnish, in order to minimize the mechanical strain introduced by sample mounting.

We first present in-field measurements taken with ambient-optics setup (Fig. 2(a)-(c)) and cryo-optics setup (Fig. 2(d)-(e)) to examine the behaviors of the MOKE susceptibility $\theta_K/B$ across $T_{CDW}$ at different magnetic fields used in prior Sagnac papers [21–23]. In both experiments, we first perform spatial scans to locate optically flat regions, and then conduct temperature sweeps at a fixed point.

With the ambient-optics setup, the Faraday background offset calibration was determined ($-112\ \mu rad/T$) [21,22] and removed from the raw signal. In Fig. 2(a) and (b), we present scanning images of $\theta_K/B$ and optical power $P_2$ in $B = 0.34\ T$ at $T = 2\ K$, procured from the same scan. There is a uniform value of $\theta_K/B \sim 24\ \mu rad/T$ in the center of optically flat regions with a good reflection ($P_2 > 2\ \mu W$). Fig. 2(c) plots MOKE susceptibility $\theta_K/B$ during temperature sweeps in $B = \pm 0.34\ T$ at the same location. They agree quantitatively with prior Irvine experiments taken at identical fields [21,22]. The most prominent feature in $\Delta\theta_K/B$ is the sharp drop of $\Delta\theta_K/B \sim -5\ \mu rad/T$ just below $T_{CDW}$ from CDW transition. Due to the absence of a Curie–Weiss shape, we attribute $\theta_K/B$ to Pauli paramagnetism, which is proportional to the density of states at the Fermi energy. The sharp drop in $\theta_K/B$ below $T_{CDW}$ can therefore be explained by a decreased density of states due to the formation of a charge density wave state below $T_{CDW}$. This observation also agrees with the reported reduction of magnetic susceptibility $\chi$ in similar samples [2].

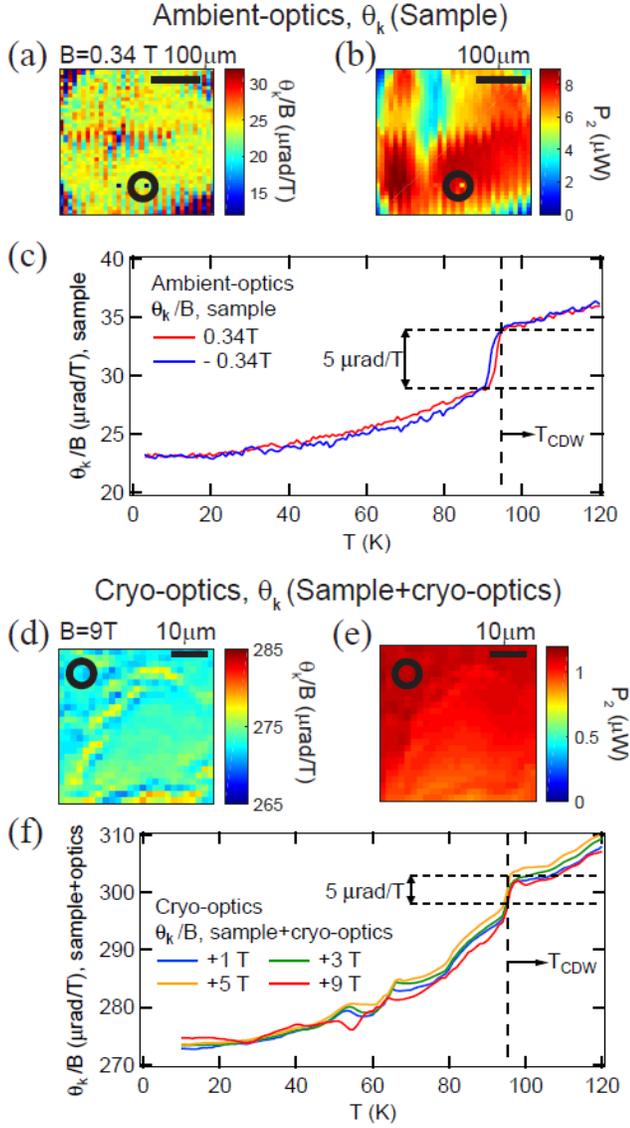

FIG. 2. In-field measurements with ambient-optics and cryo-optics setups plotted as $\theta_K/B$. (a) and (b) are scanning images of $\theta_K/B$ and optical power $P_2$ in $B = 0.34\ T$ at $T = 2\ K$, showing uniform $\theta_K$ in optically flat regions ($P_2 > 2\ \mu W$). (c) shows temperature dependence of $\theta_K/B$ in $B = \pm 0.34\ T$ at a single point as circled in (a). The Faraday contribution from ambient optics is removed. (d), (e), (f) are corresponding data taken with the cryo-optics setup at higher magnetic fields, except that the Faraday contribution from cryo-optics (250 to 280 $\mu rad/T$) is not repeatable due to thermal lag and is not subtracted. In both low (c) and high fields (f), an abrupt decrease of $\Delta\theta_K/B \sim -5\ \mu rad/T$ is evident below $T_{CDW}$.

With the cryo-optics setup, the in-field measurements are extended to higher fields as used in [23]. Since Faraday backgrounds from cryo-optics components with long thermalization time can't be calibrated reliably in temperature sweeps, they are not removed from the $\theta_K/B$ data presented in Fig. 2(d)-(e). Scanning images of $\theta_K/B$ (Fig. 2(d)) and optical power $P_2$ (Fig.2(e)) were taken with $B = 9\ T$ at $T = 2\ K$, showing a uniform value of $\theta_K/B \sim 274\ \mu rad/T$ in the scanned region. Note that this value has contributions from CsV$_3$Sb$_5$ sample and cryo-optics components combined.

Temperature sweeps of $\theta_K/B$ at a fixed location are plotted in Fig. 2(f) for $B = 1, 3, 5, 9\ T$. As expected, they are not quite repeatable because of the cryo-optics components' temperatures and hence their Faraday contributions (ranging from 250 to 280 $\mu rad/T$) do not repeat each other and display stochastic shapes due to thermal lags that are inherently stochastic. We find this non-repeatability persists even with thermal ramp rates as slow as $0.3\ K/minute$. It is therefore generally not feasible to obtain the temperature dependence of $\theta_K/B$ of the sample using the cryo-optics setup, except for a small temperature window near $T_{CDW}$ where the change of $\theta_K/B$, dubbed $\Delta\theta_K/B$, is dominated by the sample.

Closely examining the sharp changes near $T_{CDW}$ in Fig. 2(f), we found a consistent $\Delta\theta_K/B \sim -5\ \mu rad/T$ for all applied high fields $B = 1, 3, 5, 9\ T$, which fully agrees with the behaviors at $B = \pm 0.34\ T$ with the ambient-optics setup as shown in Fig. 2(c), as well as those reported in [21,22]. This value is opposite in sign to the $\Delta\theta_K/B \sim -7\ \mu rad/T$ value originally reported in the Kyoto experiment at identical high fields [23], where a sign error in that experiment has recently been acknowledged in a revised version [23].

Next, we examine the spontaneous Kerr effect during zero magnetic field warmups (ZFW) using the cryo-optics setup. The measurements were performed during ZFW after cooldowns with high magnetic fields and temperature ramping rates identical to those used in [23]. The results are presented in Fig. 3. Note that at zero field, there is no Faraday contribution from cryo-optics and hence $\theta_K$ is solely from the sample, as long as there is negligible remanent magnetic field. We have carefully verified that the remanent magnetic field in the superconducting magnet is less than ten Gauss.

A scanning image of $\theta_K$ at $B = 0$ and $T = 2\ K$ after $1\ T$ training is shown in Fig. 3(a). And the simultaneously taken optical power image is presented in Fig. 3(b). There is no observable spontaneous $\theta_K$ across the scanned region with an error bar $1\ \mu rad$. This relatively large error bar is due to the short time spent at each pixel during scanning imaging.

More accurate measurements with much smaller error bars were taken at fixed locations during very slow ZFW at 0.3 to $1\ K/minute$ temperature ramp rates, after trainings with fields of $B = +1, +3, +5, +9\ T$. The measured $\theta_K$ during these ZFW are plotted in Fig. 3(c) in their raw form without any correction of drifts. The curves are shifted with $1\ \mu rad$ vertical offsets for clarity. The overall observations of the data shown in Fig. 3(c) indicate the absence of any observable spontaneous Kerr signal at least to an uncertainty level of 0.1 $\mu rad$ which is 10 times smaller than the reported spontaneous $\theta_K$ in [23], even with similar training fields as large as 9 $T$.

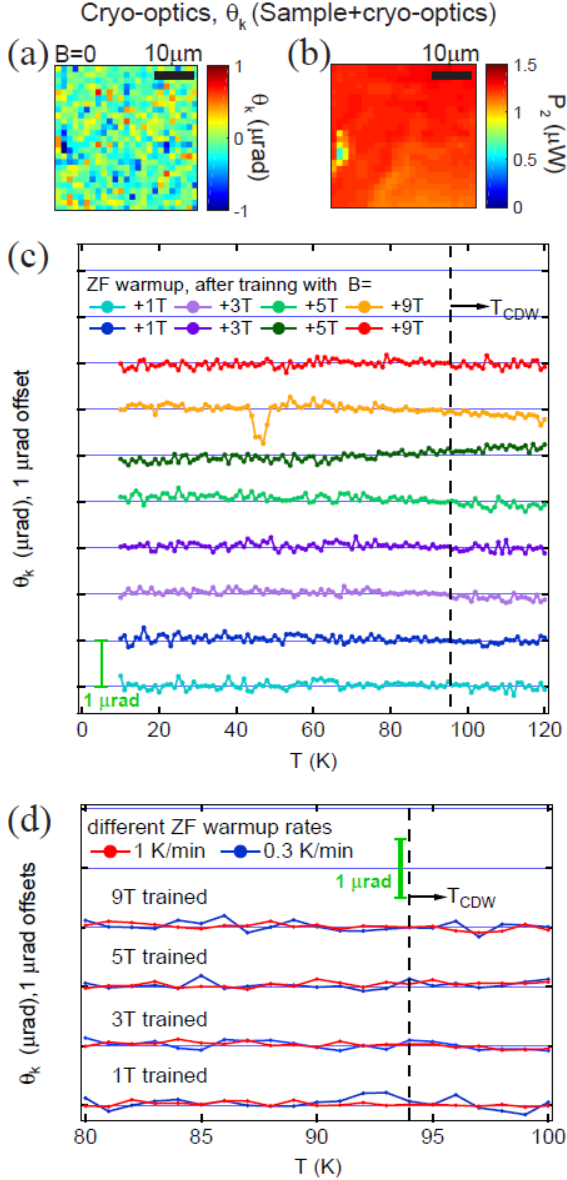

FIG. 3. Zero-field measurements with cryo-optics setup. No spontaneous $\theta_K$ was found below $T_{CDW}$ with 0.1 $\mu rad$ uncertainty. (a) and (b) are scanning images of $\theta_K$ and optical power $P_2$ at $B = 0$ and $T = 2\,K$. (c) shows $\theta_K$ during ZFW after trainings with $B = 1, 3, 5, 9\,T$ at a fixed point. (d) shows the ZFW with different thermal ramp rates of $1\,K/min$ and $0.3\,K/min$, where traces are shifted by 1 $\mu rad$ for clarity.

Here the 0.1 $\mu rad$ uncertainty level is limited by the larger-than-usual drifts as compared to the 0.03 $\mu rad$ drifts reported in faster Irvine measurements [21,22] on CsV$_3$Sb$_5$. During the very long measurement time of up to 8 hours using the very slow temperature ramp rates in [23], such drifts and even sudden changes in the optical and electronic measuring system are inevitable. As can be seen in Fig. 3(c), all 120 $K$ temperature sweeps suffered from drifts of 0.1 $\mu rad$ level. And as a demonstration of sudden change in the measurement setup, in one of the +9T trained ZFW curve (orange) between $T = 42\,K$ and $T = 50\,K$, there is an artificial drop and then a return of the detected $\theta_K$ signal of 1 $\mu rad$ as a result of powering on and off a pump that is plugged into the measurement power strip.

With unavoidable drifts in such long measurements, one must be cautious with data interpretation. While the overall observations of the data shown in Fig. 3(c) indicate the absence of an observable spontaneous Kerr signal, drifts in some measurements may mimic a change of $\theta_K$ at the 0.1 $\mu rad$ level. For example, in the two +5T trained (green and dark green) and the +9T trained (orange) data curves, the drifts around $T_{CDW}$ could be confused with an onset of a spontaneous $\theta_K$ signal of up to $\pm 0.1\,\mu rad$.

Luckily, by looking at the repeatability and sign of these "transitions", we can quickly determine that they are merely drifts instead of a true onset of the Kerr signal. The directions of these drifts are random with the same direction of training magnetic field, and they are not repeatable under identical conditions. For example, the two +5T trained measurements (green and dark green in Fig. 3(c)) drift in opposite directions, indicating 'onsets' of positive and negative spontaneous Kerr signal under the same positive training field, which is not plausible.

Temperature ramp rate is another factor that is different between experiments reported in [21,22] and in [23]. The exact CDW states in CsV$_3$Sb$_5$, such as $2 \times 2 \times 2$ or $2 \times 2 \times 4$, could in principle be affected by the speed of temperature ramping [3–6]. The Kyoto experiment [23] has used a much slower rate of 0.3 $K/min$ compared to the 1 $K/min$ rate used in the Stanford and Irvine experiments [21,22]. This difference has been suggested as a necessary condition for observing the spontaneous MOKE signal [23] (private communications). To test this hypothesis, we have performed ZFW measurements at both thermal ramp rates after training fields of $B = +1, +3, +5, +9\,T$. The results are plotted in Fig. 3(d) with 1 $\mu rad$ offsets between traces for clarity. No spontaneous $\theta_K$ is found below $T_{CDW}$ at either rate with an uncertainty of 0.1 $\mu rad$.

## IV. CONCLUTION

With comprehensive MOKE measurements using two Sagnac interferometer setups of different optical configurations with training magnetic fields of up to 9 Tesla and temperature ramp rates as slow as 0.3 $K/min$, we found no observable spontaneous MOKE signal in CsV$_3$Sb$_5$, hence no optical evidence for TRSB. This unambiguously rules out the possibility of a training-field-dependent ground state in

CsV$_3$Sb$_5$ with opposite TRSB properties. Our analyses and measurements of signal contributions from various optical components as well as signal drifts highlight the importance of proper data analysis and interpretation, especially when dealing with intricate techniques such as a Sagnac interferometer.

## V. ACKNOWLEDGMENTS

Email: xia.jing@uci.edu

This project was supported mainly by the Gordon and Betty Moore Foundation through Emergent Phenomena in Quantum Systems (EPiQS) Initiative Grant GBMF10276, and in part by NSF award DMR-1807817. S.D.W. and B.R.O. acknowledge support via the UC Santa Barbara NSF Quantum Foundry funded via the Q-AMASE-i program under award DMR-1906325.